\documentclass[11pt]{article}
\usepackage{moriond,epsfig}

\def\be{\begin{equation}}
\def\ee{\end{equation}}
\def\bea{\begin{eqnarray}}
\def\eea{\end{eqnarray}}

\def\d0{{D\O}}
\begin{document}
\vspace*{4cm}
\title{INVESTIGATIONS OF DOUBLE PARTON SCATTERING\\
EXAMPLE OF {\bf $p p \rightarrow b~\bar{b}~\rm{jet~jet}~X$}}

\author{ EDMOND L. BERGER }

\address{High Energy Physics Division, Argonne National Laboratory, Argonne, IL 60439, U.S.A}

\maketitle\abstracts{
Signature kinematic variables and characteristic concentrations in phase space of double parton scattering are discussed.   These properties should allow the double-parton contribution to $p p \rightarrow b~\bar{b}~\rm {jet~jet}~X$ at Large Hadron Collider energies to be distinguished from the usual single parton scattering contribution.  A methodology is suggested to measure the size of the double-parton cross section.}

\section{Introduction}
\begin{figure}
\begin{center}
\includegraphics[width=0.4\textwidth]{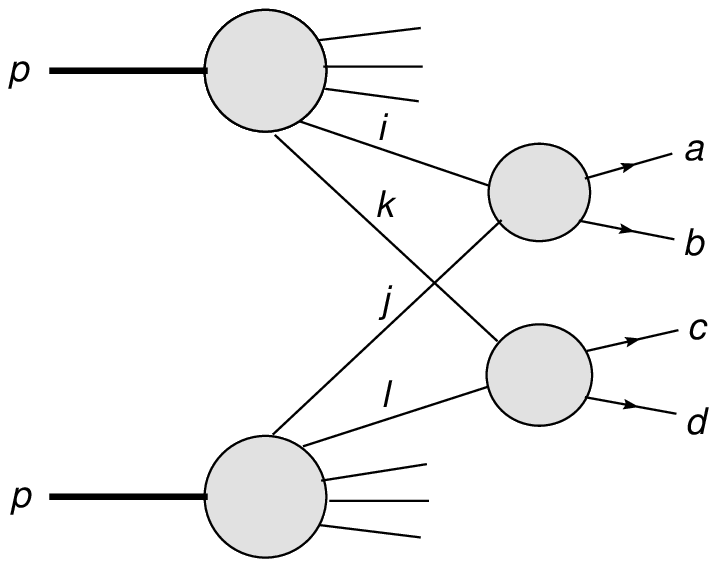}
\includegraphics[width=0.45\textwidth]{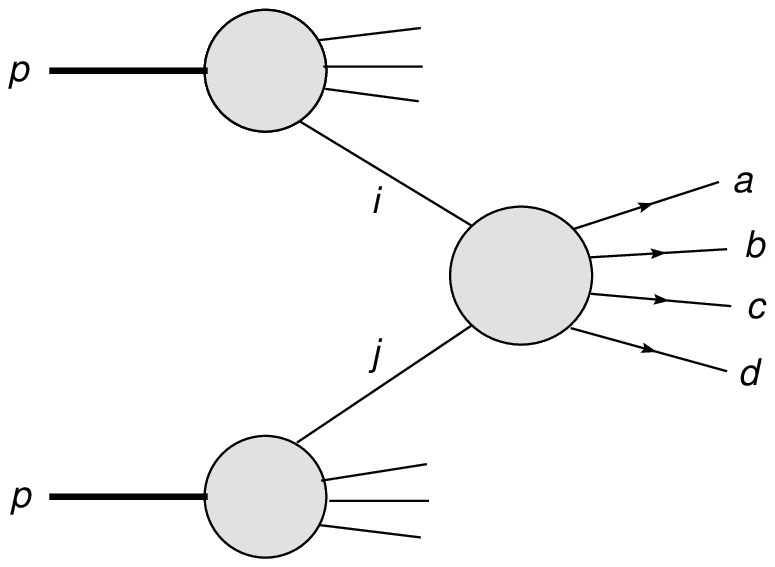}
\end{center}
\caption{(diagram on the left) Sketch of a double-parton process in which the active partons are 
$i$ and $k$ from one proton and $j$ and $l$ from the second proton.  The 
two hard scattering subprocess are $A(i~j \rightarrow a~b)$ and $B(k~l \rightarrow c~d)$. 
(diagram on the right) Sketch of a single-parton process in which the active partons are 
$i$ from one proton and $j$ from the second proton.  The 
hard scattering subprocess is $A(i~j \rightarrow a~b~c~d)$. 
\label{fig:feyn-diag}}
\end{figure}

Double parton scattering (DPS) means that two short-distance hard-scattering subprocesses occur in a given hadronic interaction, with two initial partons being active from each of the incident protons in a collision at the Large Hadron Collider (LHC).  The concept is shown for illustrative purposes in the left diagram of Fig.~\ref{fig:feyn-diag}.   It may be contrasted with conventional single parton scattering (SPS) in the right diagram, in which one short-distance subprocess occurs, with one parton active from each initial hadron.  Both contribute to the same 4 parton final state.   Processes such as sketched in the left diagram of Fig.~\ref{fig:feyn-diag} are included in descriptions of the underlying event in some Monte Carlo codes.  Our interest is to investigate whether this second hard process can be shown to be present in LHC data, as a perturbatively calculable hard part of the underlying event.  

Studies of double parton scattering have a long history theoretically, with many references to prior work listed in our paper~\cite{Berger:2009cm}, and there is evidence in data~\cite{Abazov:2009gc}.  A greater role for double-parton processes may be expected at the LHC where higher luminosities are anticipated along with the higher collision energies.  A large contribution from double parton scattering could result in a larger than otherwise predicted rate for multi-jet production, and produce relevant backgrounds in searches for signals of new phenomena.  The high energy of the LHC also provides an increased dynamic range of available phase space for detailed investigations of DPS.  
 
Our aims~\cite{Berger:2009cm} are to address whether double parton scattering can be shown to exist as a discernible contribution in well defined and accessible final states, and to establish the characteristic features that allow its measurement.  
We show that double parton scattering produces an enhancement of events in regions of phase space in which the contribution from single parton scattering is relatively small.  If such enhancements are observed experimentally, with the kinematic dependence we predict, then we will have a direct empirical means to measure the size of the double parton contribution.  In addition to its role in general LHC phenomenology, this measurement will have an impact on the development of partonic models of hadrons, since the effective cross section for double parton scattering measures the size in impact parameter space of the incident hadron's partonic hard core.  

From the perspective of sensible rates and experimental tagging, a good process to examine should be the 4 parton final state in which there are $2$ hadronic jets plus a $b$ quark and a $\bar{b}$ antiquark, {\em viz.} $b~\bar{b}~j_1~j_2$.  If the final state arises from double parton scattering, then it is plausible that one subprocess produces the $b~\bar{b}$ system and another subprocess produces the two jets.  There are, of course, many single parton scattering (2 to 4 parton) subprocesses that can result in the $b~\bar{b}~j_1~j_2$ final state, and we identify kinematic distributions that show notable separations of the two contributions.  

The state-of-the-art of calculations of single parton scattering is well developed whereas the phenomenology of double parton scattering is less advanced.  For 
$p p \rightarrow b \bar{b} j_1 j_2 X$, assuming that the two subprocesses 
$A(i~j \rightarrow b~\bar{b})$ and $B(k~l \rightarrow j_1~j_2)$ in Fig.~\ref{fig:feyn-diag} are weakly correlated, and that kinematic and dynamic correlations between the two partons from each hadron may be safely neglected, we employ the common heuristic expression for the DPS differential cross section 
\begin{eqnarray}
\label{bbjj}
d\sigma^{DPS}(p p \rightarrow b \bar{b} j_1 j_2 X) = \frac{d\sigma^{SPS}(p p \rightarrow b \bar{b} X) d\sigma^{SPS} (p p \rightarrow j_1 j_2 X)}{ \sigma_{\rm eff}}.  
\label{eq:dpscross}
\end{eqnarray}
The numerator is a product of single parton scattering cross sections.  
In the denominator, there is a term $ \sigma_{\rm eff}$ with the dimensions of a cross section.  Given that one hard scatter has taken place, $\sigma_{\rm eff}$ measures the effective probability for a second hard scatter.  Collider data~\cite{Abazov:2009gc} yield values in the range $\sigma_{\rm eff} \sim 12$~mb.  We use this value for the estimates we make, but we emphasize that the goal should be determine its value in 
experiments at LHC energies.  

The details of our calculation of the double parton and the single parton contributions to 
$p~p \rightarrow b~\bar{b}~j_1~j_2~X$ are found in our paper~\cite{Berger:2009cm}.  We perform full event simulations at the parton level and apply a series of cuts to emulate experimental analyses.  We also treat  the double parton and the single parton contributions to $4$ jet production, again finding that good separation is possible despite the combinatorial uncertainty in the pairing of 
jets~\cite{Berger:2009cm}.   

\section{Distinguishing variables}
\label{sect:variables}
\begin{figure}
\begin{center}
\includegraphics[width=0.49\textwidth]{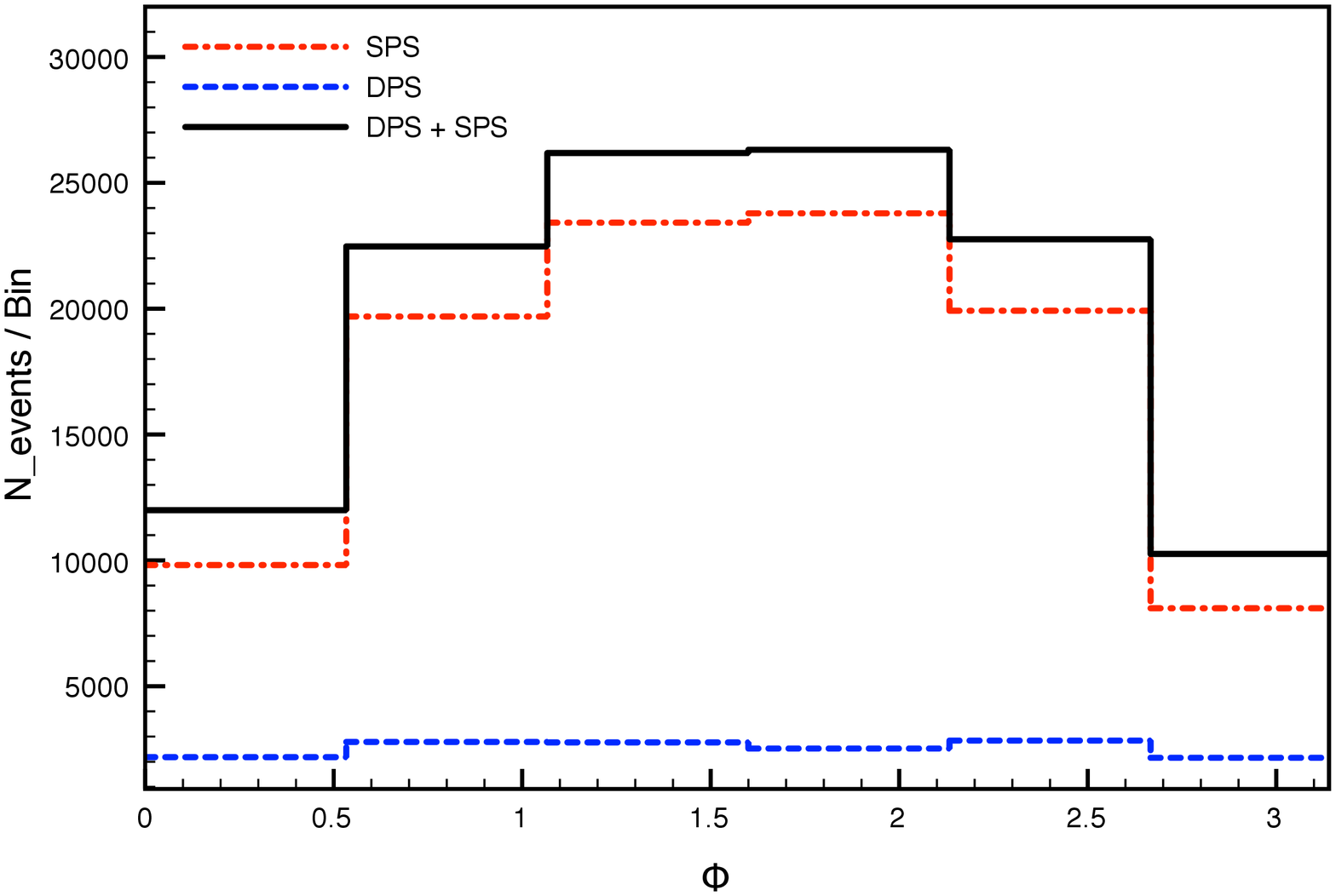}
\includegraphics[width=0.49\textwidth]{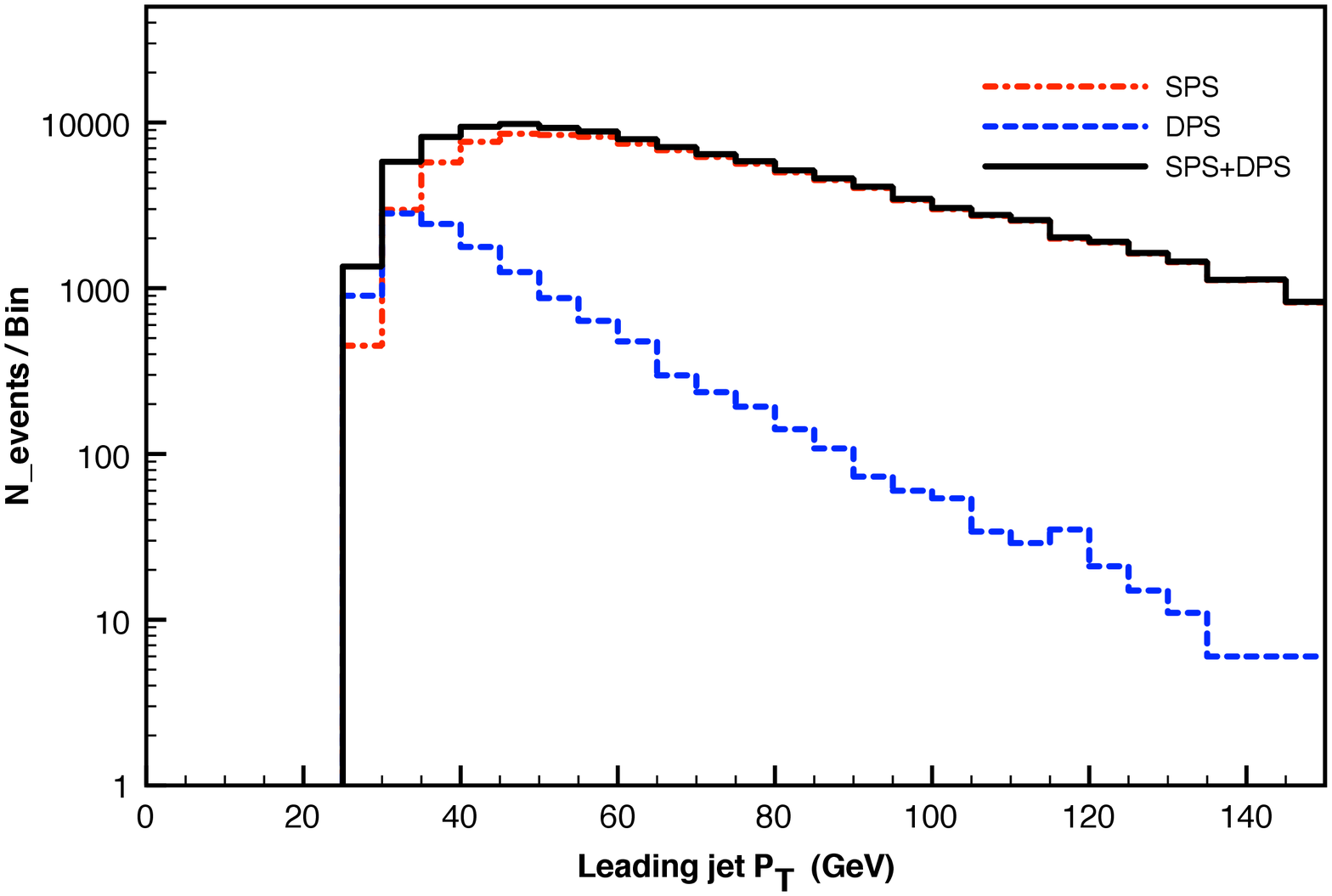}
\end{center}
\caption[]{(left panel) Event rate as a function of the angle between the two planes
  defined by the $b\bar{b}$ and $jj$ systems.  In SPS events, there is a correlation among the planes which is absent for DPS events.  
(right panel) The transverse momentum $p_T$ distribution of the leading jet in $jjb\bar b$, either a $b$ jet or a light jet $j$. }  
\label{fig:Phi-planes}
\end{figure}

Correlations in the final state are predicted to be quite different between the double parton and the single parton contributions.  For example, we examine the distribution of events as 
function of the angle $\Phi$ between the planes defined by the $b\bar{b}$ system and by the $jj$ system.  If the two scattering processes $i j \rightarrow b \bar{b}$ and $k l \rightarrow j j$ which produce the DPS final state are truly independent, one would expect to see a flat distribution in the angle $\Phi$.  By contrast, many diagrams, including some with non-trivial spin correlations, contribute to the 2 parton to 4 parton final state in SPS $i j \rightarrow b \bar{b} \rm {jet jet} $, and one would expect some correlation between the two planes. 
In the left panel of Fig.~\ref{fig:Phi-planes}, we display the number of events as a function of the angle 
between the two planes.  There is an evident correlation between the two planes in SPS, 
while the distribution is flat in DPS, consistent with the expectation that the two planes are uncorrelated.

Another dynamic difference between DPS and SPS is the behavior of event rates as a function of transverse momentum.  As an example of this, in the right panel of Fig.~\ref{fig:Phi-planes}, 
we show the transverse momentum distribution for the leading jet (either a $b$ or light $j$) for both DPS and SPS.  SPS produces a relatively hard spectrum, associated with the presence of several propagators in the hard-scattering matrix element.  On the other hand, DPS produces a much softer spectrum which (up to issues of normalization in the form of $\sigma_{\rm eff}$) can dominate at small values of transverse momentum.   For the value of $\sigma_{\rm eff}$ and the cuts that we use, SPS tends to dominate over the full range of transverse momentum considered.  The cross-over between the two contributions to the total event rate is $\sim 30$ GeV for the acceptance cuts considered.   A smaller (larger) value of $\sigma_{\rm eff}$ would move the cross-over to a larger (smaller) value of the transverse momentum of the leading jet.    

Although interesting, the two distributions in Fig.~\ref{fig:Phi-planes} would not allow the two components, DPS and SPS, to be separated.   We turn next to the search for variables that could allow 
a clear separation of the contributions.   At lowest order for a $2 \to 2$ process, the vector sum of the transverse momenta of the final state pair vanishes, although in reality, radiation and momentum mismeasurement smear the expected peak near zero.  Nevertheless, the DPS events are expected to show a reasonably well-balanced distribution in the transverse momenta of the jet pairs.  
To encapsulate this expectation for both light jet pairs and $b$-tagged pairs, we use the variable~\cite{Abazov:2009gc}:
\begin{equation}
S_{p_T}^\prime={1\over \sqrt 2} \sqrt{\left({|p_T(b_1,b_2)|\over |p_T(b_1)|+|p_T(b_2)|}\right)^2+\left({|p_T(j_1,j_2)|\over |p_T(j_1)|+|p_T(j_2)|}\right)^2}.
\label{eq:sptprime}
\end{equation}
Here $p_T(b_1,b_2)$ is the vector sum of the transverse momenta of the two final state $b$ jets, and $p_T(j_1,j_2)$ is the vector sum of the transverse momenta of the two (non $b$) jets.  We expect 
$p_T  \sim 0$ for both of these vector sums.  

\begin{figure}
\begin{center}
\includegraphics[width=0.49\textwidth]{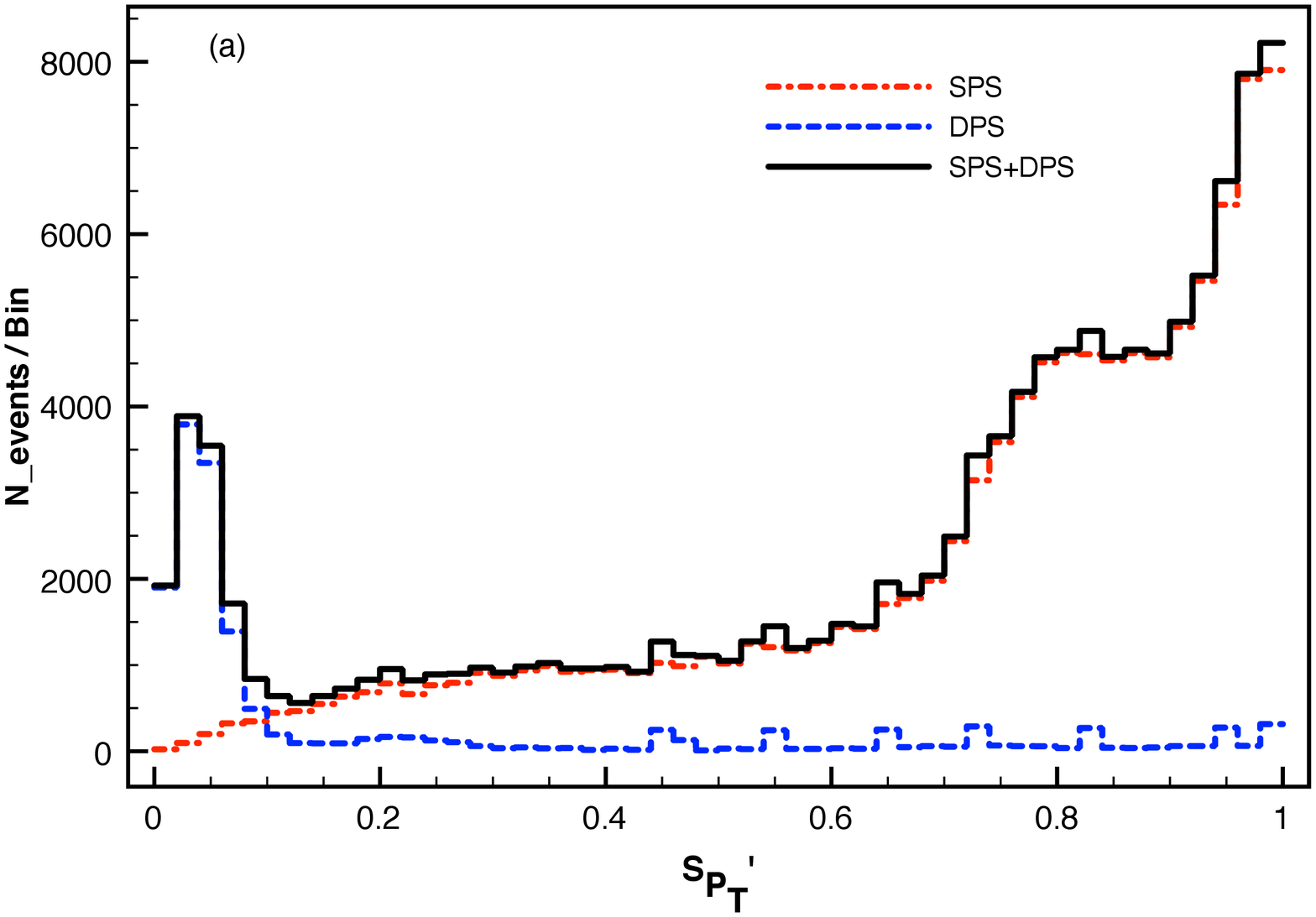}
\includegraphics[width=0.49\textwidth]{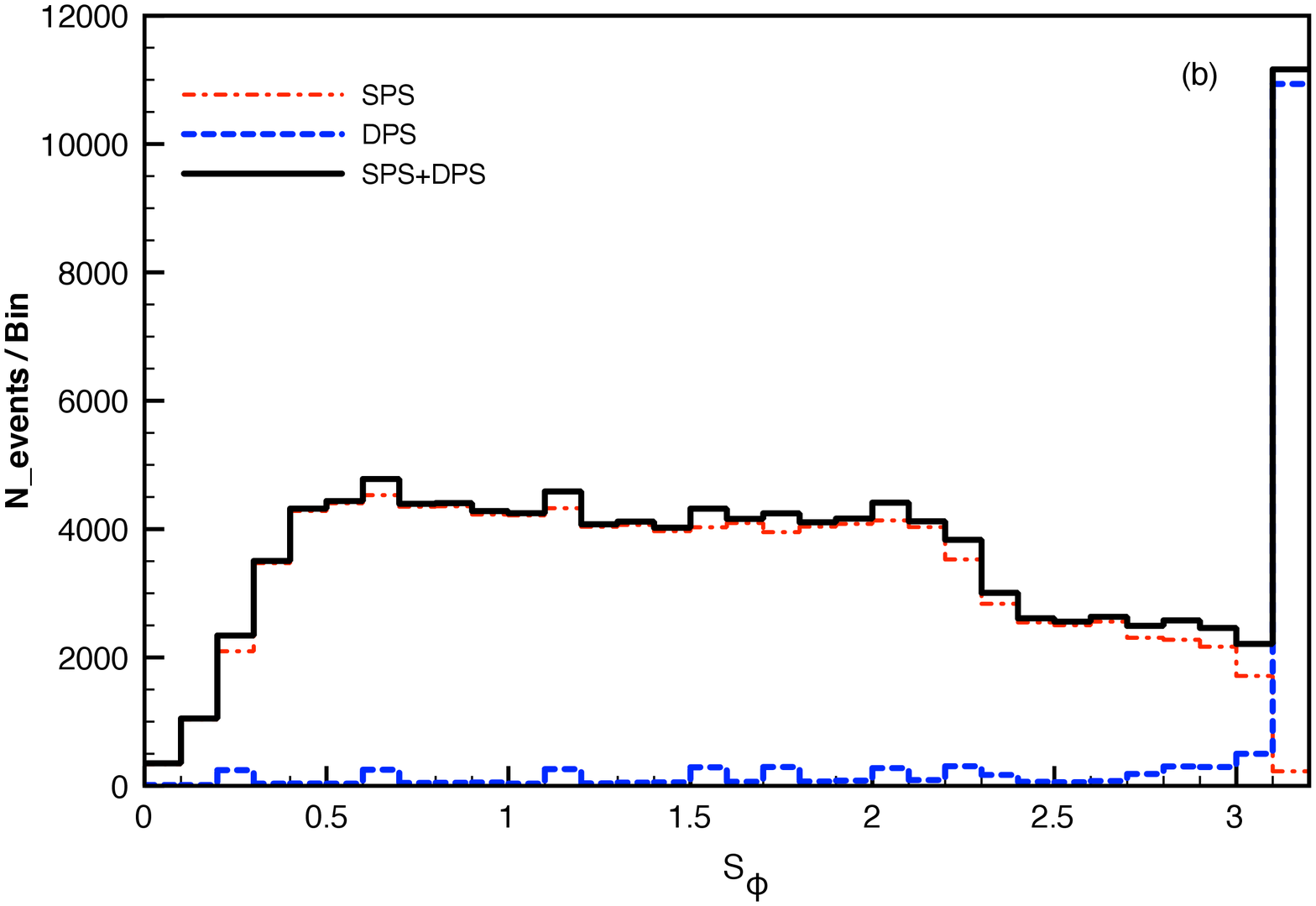}
\end{center}
\caption{ (a) Distribution of events in $S_{p_T}^\prime$ for the DPS and SPS samples.  Due to the back-to-back nature of the $2\to2$ events in DPS scattering, the transverse momenta of the jet pair and of the 
$b$-tagged jet pair are small, resulting in a small value of $S_{p_T}^\prime$.  
(b) The variable $S_\phi$ for DPS and SPS+DPS events provides a stronger separation of the underlying DPS events from the total sample when compared to $\Delta\phi$ for any pair.}
\label{fig:sptprimecut}
\end{figure}

The distribution in $S_{p_T}^\prime$ is shown in Fig.~\ref{fig:sptprimecut}(a).  As expected, the DPS events are peaked near $S_{p_T}^\prime\sim0$ and are well-separated from the total sample.  The SPS events, on the other hand, tend to be broadly distributed and show a peak near $S_{p_T}^\prime\sim1$.  The  peak near $1$ is related to the fact that a significant number of the SPS $b\bar{b}$ or $jj$ pairs arise from gluon splitting which yields a large $p_T$ imbalance and, thus, larger values of 
$S_{p_T}^\prime$.

The azimuthal angle between pairs of jets is another variable that can represent  the roughly back-to-back hard-scattering topology of the DPS events.   We expect the azimuthal angle between the pairs of jets corresponding to each hard scattering event to be strongly peaked near $\Delta \phi_{jj} \sim \Delta \phi_{bb} \sim \pi$.  Real radiation of an additional jet, where the extra jet is missed because it fails the threshold or acceptance cuts, allows smaller values of $\Delta \phi_{jj}$.  There is a clear peak near $\Delta\phi_{jj}=\pi$ for DPS events, while the events are more broadly distributed in SPS events~\cite{Berger:2009cm}.  A  secondary peak near small $\Delta\phi_{jj}$ arises from gluon splitting which typically produces nearly collinear jets.   As in the case of the $S_{p_T}^\prime$  variable, 
the separation of DPS events from SPS events becomes more pronounced if information is used from both the $b\bar{b}$ and $jj$ systems.   One distribution built from a combination of the azimuthal angle separations of both $jj$ and $b\bar b$ pairs is~\cite{Abazov:2009gc}:
\begin{equation}
S_{\phi}={1\over \sqrt 2} \sqrt{\Delta \phi(b_1,b_2)^2+\Delta \phi(j_1,j_2)^2}.
\end{equation}
The SPS events are broadly distributed across the allowed range of $S_\phi$,  shown in Fig.~\ref{fig:sptprimecut}(b).  However, the combined information from both the $b\bar{b}$ and $jj$ systems shows that the DPS events produce a sharp and substantial peak near $S_\phi \simeq \pi$ which is well-separated from the total sample.
The narrow peaks near $\Delta\phi_{jj}=\pi$ and near $S_{\phi} = 1$ are smeared somewhat once soft QCD radiation and other higher-order terms are included in the calculation.  

In our simulations, the variable $S_{p_T}^\prime$ appears to be a more effective discriminator than 
$S_{\phi}$.   However, given the leading order nature of our calculation and the absence of smearing associated with initial state soft radiation, this picture is subject to change, and a variable such as $S_\phi$ (or some other variable) may offer a clearer signal of DPS at the LHC.  Realistically, it would be valuable to study both distributions, once LHC data are available, in order to determine which is more instructive.  

\begin{figure}
\begin{center}
\includegraphics[width=0.90\textwidth]{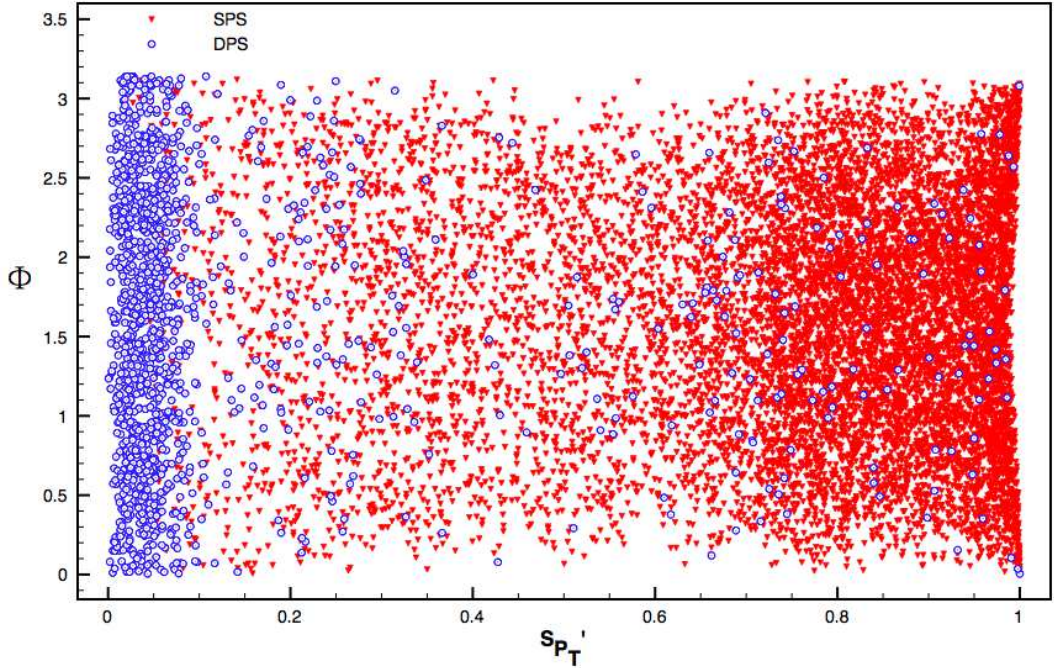}
\end{center}
\caption[]{Two-dimensional distribution of events in the inter-plane angle $\Phi$ and the scaled transverse momentum variable $S_{p_T}^\prime$ for the DPS and SPS samples.}
\label{fig:scatter1}
\end{figure}

The evidence in one-dimensional distributions for distinct regions of DPS dominance prompts the search for greater discrimination in a two dimensional distribution of one variable against another.  One scatter plot with interesting features is displayed in 
Fig.~\ref{fig:scatter1}.   The DPS events are seen to be clustered near $S'_{p_T} = 0$ and are uniformly distributed in $\Phi$.  The SPS events peak toward $S'_{p_T} = 1$ and show a roughly $\sin \Phi$ character.   While already evident in one-dimensional projections, these two  
features are more apparent in the scatter plot Fig.~\ref{fig:scatter1}.  Moreover, the 
scatter plot shows a valley of relatively low density between $S'_{p_T} \sim 0.1$ and $\sim 0.4$. 
In an experimental one-dimensional $\Phi$ distribution, one would see the sum of the DPS and SPS contributions.  If structure is seen in data similar to that shown in the scatter plot Fig.~\ref{fig:scatter1}, one could make a cut at $S'_{p_T} < 0.1$ or $0.2$ and verify whether the experimental distribution in $\Phi$ is flat as expected for DPS events.  
 
\section{Strategy and Further Work}
The clear separation of DPS from SPS events in Fig.~\ref{fig:scatter1} suggests a methodology for the study of DPS.  One can begin with a clean process such as $p p \rightarrow b \bar{b} j_1 j_2 X$ and examine the distribution of events in the plane defined by  $S'_{p_T} $ and $\Phi$.  We expect to see a concentration of events near $S'_{p_T} = 0$ that is uniformly distributed in $\Phi$.  These are the DPS events.  Assuming that a valley of low density is observed between $S'_{p_T}  \sim 0.1 $ and $\sim 0.4$, one can make a cut there that produces an enhanced DPS sample.  Relative to the overall sample, this enhanced sample should show a more rapid decrease of the cross section as a function of the transverse momentum of the leading jet, and the DPS enhanced sample can be used to measure 
$\sigma_{\rm eff}$.  A similar examination of other final states, such as 4 jet production, will answer whether the extracted values of $ \sigma_{\rm eff}$ are roughly the same.  Theoretical and experimental studies of other processes can follow, such as $b \bar{b} t \bar{t}$, $W {\rm j j}$, and $H {\rm j j}$.  

On the phenomenological front, next-to-leading order (NLO) expressions should be included for both the SPS and DPS contributions.  The NLO effects are expected to change normalizations and, more importantly, the distributions in phase space.  The sharp peaks near $S_\phi \simeq \pi$ in 
Fig.~\ref{fig:sptprimecut}(b) and $S_{p_T}^\prime = 0$ in Fig.~\ref{fig:sptprimecut}(a) will be broader and likely displaced somewhat.  The weak correlation between subprocesses assumed in 
Eq.~(\ref{eq:dpscross}) cannot be strictly true~\cite{Gaunt:2009re}.  With a large enough data sample at the LHC one could investigate the extent to which correlations play a significant role.  

Finally, it would be good to examine the theoretical underpinnings of Eq.~(\ref{eq:dpscross}) and, in the process, gain better insight into the significance of $\sigma_{\rm eff}$.   A firm basis is 
desirable for Eq.~(\ref{eq:dpscross}) starting from the formal expression for the differential cross section in terms of the absolute square of the full matrix element integrated over phase space:
\begin{eqnarray}
d\sigma(p p \rightarrow b \bar{b} j_1 j_2 X) = \frac{1}{2s} | {M (p p \rightarrow b \bar{b} j_1 j_2 X)}|^2 d PS_{b \bar{b} j_1 j_2 X}.  
\end{eqnarray}

The amplitude $M (p p \rightarrow b \bar{b} j_1 j_2 X)$ should include a sum of amplitudes for 
2-parton collisions (one active from each incident hadron, i.e., $2 \rightarrow 4$);  
3-parton collisions (two active from one hadron and one active from the other); and 
4-parton collisions (two active from each hadron or three from one and one from the other), 
and so forth that all yield the same 4 parton final state.  There will be contributions to the final state 
from the squares of individual amplitudes as well as interference terms.  
Specializing to $4 \rightarrow 4$, the DPS case, one would start from a 4-parton $\rightarrow$ 4-parton hard part.   Not evident at this time is how the four-parton matrix element can be reduced to a product 
of two matrix elements for the single parton scatterings, needed for  Eq.~(\ref{eq:dpscross}).  The 
demonstration of clear DPS signals in LHC data would be an important stimulus for further theoretical studies.  

\section*{Acknowledgments}
This work was done in collaboration with Chris Jackson and Gabe Shaughnessy and supported financially by the U.~S.\ Department of Energy under Contract No.\ DE-AC02-06CH11357.   I acknowledge valuable discussions with Tom LeCompte and Jianwei Qiu.

\end{document}